\documentclass{ws-ijmpd}
\usepackage[super,compress]{cite}
\usepackage{graphicx}
\usepackage{amssymb,amsmath,accents}
\usepackage{amscd,amsfonts}
\usepackage{mathrsfs}
\usepackage{hyperref}
\allowdisplaybreaks
\begin{document}

\markboth{Muhammad Usman and Asghar Qadir}
{Dark energy from two Higgs doublet model}

%
\catchline{}{}{}{}{}
%

\title{
The extra scalar degrees of freedom from the two Higgs doublet model for dark energy	
}

\author{Muhammad Usman\footnote{muhammad\_usman\_sharif@yahoo.com}} \author{\space and Asghar Qadir\footnote{asgharqadir46@gmail.com}}
\address{Department of Physics, \\ School of Natural Sciences (SNS),\\
	National University of Sciences and Technology (NUST),\\
	Sector H-12, Islamabad 44000, Pakistan}

\maketitle

\begin{history}
\received{Day Month Year}
\revised{Day Month Year}
\end{history}

\begin{abstract}

In principle a minimal extension of the standard model of Particle Physics, the two Higgs doublet model, can be invoked to explain the scalar field responsible of dark energy. The two doublets are in general mixed. After diagonalization, the lightest CP-even Higgs and CP-odd Higgs are jointly taken to be the dark energy candidate. 
The dark energy obtained from Higgs fields in this case is indistinguishable from the cosmological constant.
\end{abstract}


\ccode{PACS numbers: 95.36.+x; 12.60.Fr}

\section{Introduction}\label{sec:intro}
There are at least three components which contribute to the total energy
density of the Universe: (a) non-relativistic matter; (b) relativistic
matter; (c) dark energy. Dark energy, which constitutes about $70\%$ of the
present energy density of the Universe, is taken to be causing the current
accelerated expansion of the Universe. Several possible candidates of dark energy can be found in the literature, such as the cosmological
constant ($\Lambda$) \cite{Peebles,muhammadsami}; scalar field models (e.g.
quintessence, $K$-essence, tachyon field, phantom (ghost) field
\cite{trodden,0264-9381-19-17-311,PhysRevD.70.107301}, dilatonic dark energy,
Chaplygin gas) \cite{muhammadsami} and vector fields\cite{PhysRevD.78.063005,Jiménez2010175,1475-7516-2009-03-016,Böhmer2007,1475-7516-2004-07-007,vectorinflation}. Modified gravity has been proposed in order to explain the apparent accelerated expansion of the Universe \cite{Faraoni,Nojiri}.

Apart from the cosmological constant, which automatically arises in classical relativity as a constant of integration in obtaining the Einstein field equations, the other proposals go beyond all known Physics of gravity and/or the standard model of Particle Physics (SM). The most discussed one is scalar fields which were first used by Alan Guth \cite{guth} and Andre Linde \cite{Linde} to propose inflation \cite{Bassett} as a possible solution to the horizon and flatness problems. 
As such all the proposals seem to be simply speculations hoping to hit on the correct explanation. Given that there would, in principle, be infinitely many
possible suggestions, the chances of hitting on the correct one by chance seems to us to be negligibly small. It seems more reasonable to try to stick as closely to known Physics as possible. One way would be to accept that there is a small cosmological constant. In this paper we make the point of principle that we {\it can} manage an explanation while barely going beyond the standard model of Particle Physics. While the present proposal does not make a significant prediction, it does prove that this approach can
work and it may be worthwhile to search for a predictive proposal on these lines.

The essential idea is that the standard model of Particle Physics has no
restriction on the Higgs sector, either with reference to the number of Higgs
fields or to their masses. 
For definiteness in the minimal standard model only one Higgs doublet is considered. This minimal model has withstood all laboratory test for four decades. However, there is no reason to believe that no other Higgs fields can become relevant.
As such, one might look for a second Higgs that will provide late time inflation and is not inconsistent with observations. We have found that this idea does work in the minimal extension of the SM 
which is the two Higgs doublet model (2HDM). Though a prediction emerges from here, unfortunately one can not expect to test it in the foreseeable future.


The homogeneous, isotropic Universe model is described by the Friedmann-Robertson-Walker (FRW) metric and its dynamics is described by the Friedmann equations, which are
\begin{eqnarray}
H^2 &=& \dfrac{1}{3}\rho-\dfrac{k}{a^2}~, \label{1stFriedmannequation}
\\
\dfrac{\ddot{a}}{a} &=& -\dfrac{1}{6}\left(1+3\omega_{\text{eff}}\right)\rho~, \label{2ndFriemannequation}
\end{eqnarray}
where $a$ is the scale factor, $H=\dot{a}/a$ is the Hubble parameter, $\rho$ is the total energy density, $k$ is the spatial curvature, $\omega_{\text{eff}}=\Omega_{DE}\text{\space}\omega_{DE}+\Omega_{R}\text{\space}\omega_{R}+\Omega_{M}\text{\space}\omega_{M}$ is the effective equation of state (EoS). In deriving the above equations barotropic equation of state $P=\omega\rho$ has been used along with the Planck units, $\hbar=c=1$ and $(8\pi G)^{-1/2}=M_P=1$, which will also be used in all subsequent equations, $M_P$ is the reduced Planck mass here. From eq. (\ref{2ndFriemannequation}), we see that $\ddot{a}>0$ when $\omega_{\text{eff}}<-\frac{1}{3}$.

Using the law of conservation of energy, it can easily be derived that
\begin{eqnarray}\label{conservationofenergy}
\rho\propto a^{-3(1+\omega)}~.
\end{eqnarray}
From eq. (\ref{conservationofenergy}), scalar fields are put into two categories on the basis of their EoS parameter $\omega$:
\emph{Quintessence} fields ($-1<\omega<-1/3$); and \emph{Phantom} fields ($\omega<-1$) \cite{muhammadsami}. 
When $\omega=-1$ the energy density does not change as the Universe expands. 

Assuming that the present Universe can be described by the 2HDM, we assume that both the phase transitions corresponding to the doublets have occurred. The case when the phase transition in one doublet has occurred and the second doublet has zero vacuum expectation value (VeV) is called the inert doublet model. If the second doublet does not couple to the fermions via Yukawa interactions, the Higgs fields of the second doublet can be considered as a dark energy candidate when the Universe is described by the inert doublet model \cite{Usman}.
Previously, Greenwood et al \cite{PhysRevD.79.103003} considered the appearance of a second vacuum in the SM Higgs potential. They showed that if the second order electroweak phase transition is followed by a first order phase transition (which they were able to achieve for a certain range of parameters of their model) then this phase transition can cause the required late time acceleration of the Universe. Carroll et al \cite{trodden}
have also given an independent phantom field model where they have also discussed the stability of energy density along with the phantom field as a possible dark energy candidate. 
Onemli and Woodard \cite{0264-9381-19-17-311,PhysRevD.70.107301} have also showed that quantum effects in Cosmology could cause a violation of the null and weak energy conditions (i.e. $\rho+P<0$ in their model) causing $\omega<-1$ on cosmological scales without introducing any ghost, phantom, etc.

The paper is organized as follows; in the next section we review the 2HDM and obtain constraints on the parameters of the 2HDM's Higgs potential to be positive along with the calculations of the minima of the potential. In section 3, phase transitions in the 2HDM is discussed. Section 4 is devoted to determining whether the Higgs field(s) can be a possible candidate for dark energy. Section 5 gives a conclusion and discussion.
\section{The two Higgs doublet model (2HDM)}
The electroweak symmetry in the SM is broken spontaneously by the non-zero VeV of the Higgs field(s) via Higgs mechanism. The Lagrangian which describes any model in Particle Physics is
\begin{equation}\label{L}
\mathscr{L}=\mathscr{L}^{SM}_{gf}+\mathscr{L}_{Y}+\mathscr{L}_{Higgs}~.
\end{equation}
Here, $\mathscr{L}^{SM}_{gf}$ is the $SU_{C}(3){\otimes}SU_{L}(2){\otimes}U_{Y}(1)$ is the SM interaction of the fermions and gauge bosons (force carriers)\footnote{
	\begin{eqnarray}\begin{array}{rcl}\label{Lsm}
	\mathscr{L}^{SM}_{gf} &=& -\frac{1}{4}G_{\mu \nu}G^{\mu \nu}-\frac{1}{4}W_{\mu \nu}W^{\mu \nu}-\frac{1}{4}B_{\mu \nu}B^{\mu \nu}
	\\ & &
	+{\bar{\psi}}_{L}^{i}\dot{\iota}{\gamma^{\mu}}{{\nabla}_{\mu}^{EW}}{{\psi}_{L}^{i}}+{\bar{\psi}}_{R}^{i}{\iota}{\sigma^{\mu}}{{\nabla}_{\mu}^{EW}}{{\psi}_{R}^{i}}
	\\ & &
	+{\bar{\chi}}_{L}^{i}{\iota}{\gamma^{\mu}}{{\nabla}_{\mu}^{SM}}{{\chi}_{L}^{i}}+{\bar{U}}_{R}^{i}{\iota}{\sigma^{\mu}}{{\nabla}_{\mu}^{SM}}{{U}_{R}^{i}}
	\\ & & +{\bar{D}}_{R}^{i}{\iota}{\sigma^{\mu}}{{\nabla}_{\mu}^{SM}}{{D}_{R}^{i}}~. \nonumber
	\end{array}\end{eqnarray}}
, $\mathscr{L}_{Y}$ is the Yukawa interaction of fermions with the Higgs field(s)\footnote{
	\begin{equation*}\begin{array}{rcl}\label{Lyukawa}
	\mathscr{L}_{Y} &=& -Y_{ij}^u{\bar{\chi}}_L^i\tilde{\phi_1}U_R^j-Y_{ij}^d{\bar{\chi}}_L^i{\phi_1}D_R^j-Y_{ij}^e{\bar{\psi}}_L^i{\phi_1}\psi_R^j-\text{h.c.} ~,
	\end{array}\end{equation*}
	where $\psi_L^i$ are the left handed leptons doublets, $\psi_R^i$ are the right handed leptons singlets, $\chi_L^i$ are the left handed quark doublets, $U_R^i$ and $D_R^i$ are the right handed quark singlets. $i$ runs from 1-3. $\phi_1$ is the SM like Higgs doublet.
}
and $\mathscr{L}_{Higgs}$ is the Higgs field Lagrangian where
\begin{equation}\label{LH}
\mathscr{L}_{Higgs}=T_{H}-V_{H}~,
\end{equation}
$T_{H}$ being the kinetic term of the Higgs field(s) and $V_{H}$ the potential of the Higgs field(s)\footnote{In the SM one scalar isodoublet with hypercharge $Y=1$ is sufficient to make the theory complete and gauge invariant and hence in SM
	\begin{equation*}\begin{array}{rcl}\label{LSMHiggs}
	\mathscr{L}_{Higgs}=T_{H}-V_{H}=(D_{\mu}\phi)^{\dagger}(D^{\mu}\phi)-(-\frac{{\mu}^{2}}{2!}{\phi}^2+\frac{\lambda}{4!}{\phi}^4)
	\end{array}\end{equation*}
	$$\phi=\begin{pmatrix}
	\phi^+ \\ \phi^0
	\end{pmatrix} \text{\qquad\qquad and\qquad\qquad} \tilde{\phi}=\dot{\iota}\sigma^2\bar{\phi}=\begin{pmatrix}
	\bar{\phi}^0 \\ -\bar{\phi}^+
	\end{pmatrix}.$$}.
Here,
\begin{align}\begin{split}\label{TH}
T_H=&({D_1}_{\mu}\phi_1)^\dagger ({D_1}^{\mu}\phi_1) + ({D_2}_{\mu}\phi_2)^\dagger ({D_2}^{\mu}\phi_2) \\ &+\left [ {\chi}({D_1}_{\mu}{\phi}_{1})^{\dagger}({D_2}^{\mu}{\phi}_{2})+{\chi^{*}}({D_2}_{\mu}{\phi}_{2})^{\dagger}({D_1}^{\mu}{\phi}_{1}) \right ],
\end{split}\end{align}
and
\begin{align}
V_H &=\text{\space} V_1+V_2+V_{int} ~, \label{V12int}\\
\begin{split}\label{VH}
&=\text{\space}\rho_1\exp(\Lambda_1 m_{11}^2\phi_1^\dagger\phi_1)+\rho_3\exp{\Big(}\dfrac{1}{2}\Lambda_3\lambda_1(\phi_1^\dagger\phi_1)^2{\Big)} \\ &\text{\space\space\space} +\rho_2\exp(\Lambda_2 m_{22}^2\phi_2^\dagger\phi_2)+\rho_4\exp{\Big(}\dfrac{1}{2}\Lambda_4\lambda_2(\phi_2^\dagger\phi_2)^2{\Big)} \\ &\text{\space\space\space} +{m_{12}^2}(\phi_1^\dagger \phi_2)+{{m_{12}^2}^*}(\phi_2^\dagger \phi_1)
+\lambda_3 (\phi_1^\dagger\phi_1)(\phi_2^\dagger\phi_2)
\\ &\text{\space\space\space}
+\lambda_4(\phi_1^\dagger\phi_2)(\phi_2^\dagger\phi_1)+\dfrac{1}{2}{\big[}\lambda_5(\phi_1^\dagger\phi_2)^2 +\lambda_5^* (\phi_2^\dagger\phi_1)^2{\big]} \\ &\text{\space\space\space}
+\lambda_6(\phi_1^\dagger\phi_1)(\phi_1^\dagger\phi_2)
+\lambda_6^*(\phi_1^\dagger\phi_1)(\phi_2^\dagger\phi_1) \\ &\text{\space\space\space} +\lambda_7(\phi_2^\dagger\phi_2)(\phi_1^\dagger\phi_2)+\lambda_7^*(\phi_2^\dagger\phi_2)(\phi_2^\dagger\phi_1)~,
\end{split}
\end{align}
where
$V_1$ and $V_2$ in eq. (\ref{V12int}) are the Lagrangian of Higgs field $\phi_1$ (given by first two terms of RHS of eq. (\ref{VH})) and $\phi_2$ (given by $3^{rd}$ and $4^{th}$ terms of RHS of eq. (\ref{VH})) respectively, $V_{int}$ is the interaction Lagrangian of fields $\phi_1$ and $\phi_2$, given by the remaining terms, and ${D_i}_\mu$ ($i=1,2$) are the covariant derivatives given by
$${D_1}_\mu=\partial_\mu+\dot{\iota}\dfrac{g_1}{2}\sigma_i {W^i}_\mu+\dot{\iota}\dfrac{g_1'}{2}B_\mu ~,\text{\qquad}
{D_2}_\mu=\partial_\mu+\dot{\iota}\dfrac{g_2}{2}\sigma_i {W^i}_\mu+\dot{\iota}\dfrac{g_2'}{2}B_\mu ~,$$
$$
\phi_{i}=
\begin{bmatrix}
\phi^{+}_{i} \\
\dfrac{1}{\sqrt{2}}\left(\eta_i + \dot{\iota}\chi_i +\nu_i\right) \\
\end{bmatrix}~,\text{\qquad}
\phi_{i}^{\dagger}=
\begin{bmatrix}
\phi^{-}_{i} & \dfrac{1}{\sqrt{2}}\left(\eta_i - \dot{\iota}\chi_i +\nu_i\right)
\end{bmatrix}.
$$
The dimensions of different quantities are
$[\rho_i]^{-1}=[\Lambda_i]=[L]^4,\text{\space} [m_{ii}^2]=[L]^{-2},\text{\space} [\phi_i]=[L]^{-1}$
and $[\lambda_i]=[L]^0$. Here, ``$L$'' denotes the length dimension. The fields $\phi^{+}_{i}$, $\phi^{-}_{i}$, $\eta_i$ and $\chi_i$ are the hermitian Higgs fields, $\phi^{\pm}_{i}$ are charged whereas other fields are neutral also $\eta_i$ is the CP-even fields where as $\chi_i$ is the CP-odd field, $\nu_i$ is the VeV of the doublet $\phi_i$. The VeV of the Higgs' should satisfy the relation $$\nu_1^2+\nu_2^2=\nu^2=1/\sqrt{2}G_F\approxeq(246\text{GeV})^2.$$

The Higgs doublets $\phi_1$ and $\phi_2$ have a mixing angle $\beta$, where $\beta=\tan(\nu_2/\nu_1)$ in such a way that one of the new Higgs doublets acts as an identical way as the SM Higgs doublet while the other new Higgs doublet is an extra parameter in the theory. The new Higgs then can be taken as the candidate to explain the phenomena which are not compatible with the SM predictions. Thus, the new Higgs doublets are
\begin{equation}\label{newH1}
\begin{array}{rcl}
H_1 = \cos\beta~\phi_1+\sin\beta~\phi_2 =
\begin{bmatrix}
G^{+}_1 \\
\dfrac{1}{\sqrt{2}}\left(h_1+ \dot{\iota}G^{0}+\nu\right) \\
\end{bmatrix},
\end{array}
\end{equation}
\begin{equation}\label{newH2}
\begin{array}{rcl}
H_2 = -\sin\beta~\phi_1+\cos\beta~\phi_2 =
\begin{bmatrix}
-H^{+} \\
\dfrac{1}{\sqrt{2}}\left(h_2 + \dot{\iota}A\right) \\
\end{bmatrix},
\end{array}
\end{equation}
Here, $G$ represents the Goldstone field while the other symbols represent normal field(s).
The mass matrix for CP-even Higgs fields $h_1$ and $h_2$ is not diagonalized by the rotation of doublets through $\beta$. Thus, another rotation of new Higgs fields $h_1$ and $h_2$ is performed (say by angle $\gamma$) to make the mass matrix of the new CP-even fields diagonal. For simplicity, we can say that perform the rotation by the angle $\alpha$ of the CP-even Higgs fields $\eta_1$ and $\eta_2$ such that the mass matrix of the CP-even Higgs fields is diagonalized ($\alpha$ here then will be a combination of $\beta$ and $\gamma$). Now, from eq. (\ref{VH})
\begin{equation}\label{mA}
m_A^2=\dfrac{m_{12}^2}{\sin\beta\cos\beta}-\dfrac{\nu^2}{2}\left(2\lambda_5+\lambda_6\cot\beta+\lambda_7\tan\beta\right),
\end{equation}
\begin{align}\begin{split}\label{mh+}
m_{H^{+}}^2=&\text{\space\space}\dfrac{m_{12}^2}{\sin\beta\cos\beta} 
-\dfrac{\nu^2}{2}\left(\lambda_4+\lambda_5+\lambda_6\cot\beta+\lambda_7\tan\beta\right) \\ &\hspace{-0.3cm} =m_A^2+\dfrac{\nu^2}{2}\left(-\lambda_4+\lambda_5\right).
\end{split}\end{align}
The mass matrix of CP-even Higgs $\eta_1$ and $\eta_2$ is given by
\begin{equation}\label{meta1eta2}
\mathcal{M}^2=	m_A^2 \begin{bmatrix}
\sin^2\beta && -\sin\beta\cos\beta \\
-\sin\beta\cos\beta && \cos^2\beta
\end{bmatrix}
+\nu^2\mathcal{B}^2,
\end{equation}
where
\begin{align}
\hspace{-0.03cm} \mathcal{B}^2= \Bigg[
\begin{split}
&\lambda_1^{'}\cos^2\beta+2\lambda_6\sin\beta\cos\beta+\lambda_5\sin^2\beta \\ & \left(\lambda_3+\lambda_4\right)\sin\beta\cos\beta+\lambda_6\cos^2\beta+\lambda_7\sin^2\beta
\end{split} \nonumber
\\ \nonumber ~ \\
\begin{split}
& \hspace{-0.7cm} \text{\qquad\qquad\qquad\qquad}\left(\lambda_3+\lambda_4\right)\sin\beta\cos\beta+\lambda_6\cos^2\beta+\lambda_7\sin^2\beta \\ & \hspace{-0.7cm} \text{\qquad\qquad\qquad\qquad}\lambda_2^{'}\sin^2\beta+2\lambda_7\sin\beta\cos\beta+\lambda_5\cos^2\beta
\end{split} \Bigg],
\end{align}
\noindent $\lambda_1^{'}=\dfrac{1}{2}\left(\rho_3\Lambda_3\lambda_1+\rho_1(\Lambda_1 m_{11}^2)^2\right) \text{\space and \space} \lambda_2^{'}=\dfrac{1}{2}\left(\rho_4\Lambda_4\lambda_2+\rho_2(\Lambda_2 m_{22}^2)^2\right).$

With the rotation of fields $\eta_1$ and $\eta_2$ by an angle $\alpha$, the masses of the CP-even states can be obtained as \cite{PhysRevD.67.075019}
\begin{equation}\label{mCP}
\begin{bmatrix}
m_H^2 && 0 \\
0 && m_h^2
\end{bmatrix} = \mathcal{R}(\alpha)\mathcal{M}^2\mathcal{R}^T(\alpha)~,
\end{equation}
where $m_H \geq m_h$. The masses are explicitly given as

\begin{align}\begin{split}\label{mHh}
m_{H,h}^2=\dfrac{1}{2}\Bigg(
\mathcal{M}_{11}^2 & +\mathcal{M}_{22}^2 
\pm\sqrt{\left(\mathcal{M}_{11}^2-\mathcal{M}_{22}^2\right)^2+4\left(\mathcal{M}_{12}^2\right)^2}\Bigg).
\end{split}\end{align}
The Higgs vacuum energy of the potential given by eq. (\ref{VH}) is
\begin{align}
\begin{split}\label{Evac}
E_{vac}&=\rho_1 \exp\left(\dfrac{1}{2}m_{11}^2\Lambda_1\nu_1^2\right)+\rho_2 \exp\left(\dfrac{1}{2}m_{22}^2 \Lambda_2\nu_2^2\right) \\ & \text{\quad}+\rho_3 \exp\left(\dfrac{1}{8}\lambda_1\Lambda_3\nu_1^4\right)+\rho_4 \exp\left(\dfrac{1}{8}\lambda_2\Lambda_4\nu_2^4\right) \\ & \text{\quad}+ \dfrac{1}{4}\lambda_3\nu_1^2\nu_2^2+\dfrac{1}{4}\lambda_4\nu_1^2\nu_2^2+\dfrac{1}{8}\left(\lambda_5+\lambda_5^*\right)\nu_1^2\nu_2^2 \\ & \text{\quad}+ \dfrac{1}{4}\left(\lambda _6+\lambda _6^*\right)\nu _1^3\nu _2+ \dfrac{1}{4} \left(\lambda_7+\lambda_7^*\right) \nu_1 \nu_2^3.
\end{split}
\end{align}

Since we are looking to provide the accelerated expansion to the Universe from Higgs field(s), the candidate Higgs must have their life time longer than or equal to the current age fo the Universe (i.e. a stable Higgs is required for the scalar field dark energy model). This can be achieved by imposing a discrete $\mathbb{Z}_2$ symmetry $\phi\rightarrow-\phi$. There are two types of $\mathbb{Z}_2$ symmetry breakings: soft and hard. When $\mathbb{Z}_2$ symmetry is broken by $(\phi_i^\dagger\phi_j)$ type terms then it is said to be softly broken and when $\mathbb{Z}_2$ symmetry is broken by $(\phi_i^\dagger\phi_j)(\phi_k^\dagger\phi_l)$ type terms then symmetry is said to be hardly broken.
The terms of the Higgs potential (given by eq. (\ref{VH})) containing $m_{12}^{2}$ describe the soft symmetry breaking of $Z_{2}$ symmetry, whereas the terms $\lambda_{6}$ and $\lambda_{7}$ describe the hard symmetry breaking of $\mathbb{Z}_2$ symmetry. In the absence of these terms along with no cross kinetic term ($\chi=0$), the 2HDM's Higgs Lagrangian has a perfect $Z_{2}$ symmetry \cite{Ginzburg-2}. There are two $\mathbb{Z}_2$ symmetries corresponding to the doublets of the 2HDM:
\begin{eqnarray}
\text{I:\qquad\qquad}\phi_1 \longrightarrow &-\phi_1~, \text{\qquad\qquad} \phi_2 \longrightarrow &\text{\space\space}\phi_2~, \label{1stZ2symmetry}
\\
\text{II:\qquad\qquad}\phi_1 \longrightarrow &\text{\space\space}\phi_1~, \text{\qquad\qquad} \phi_2 \longrightarrow &-\phi_2~.	\label{2ndZ2symmetry}
\end{eqnarray}

It should also be noted that even if the Higgs field(s) has(have) decayed, vacuum energy given by eq. (\ref{Evac}) would still be able to provide accelerated expansion provided it dominates at some time in the life span of the Universe.

The stability of the vacuum of the Higgs potential (given by eq. (\ref{VH})) implies that the potential must be positive for all asymptotically large values of the Higgs field. Thus, the positivity of the Higgs potential implies \cite{2hdmc,kaffaskhater,Usman}
\begin{eqnarray}\label{positivitycontraint1}
\lambda_1^{'},\lambda_2^{'}>0~, \text{\qquad}\lambda_3>-2~\sqrt{\lambda_1^{'}\lambda_2^{'}}~,
\end{eqnarray}
when $\lambda_6=\lambda_7=0$, then we also have
\begin{eqnarray}\label{positivitycontraint2}
\lambda_3+\text{min}[0,\lambda_4-\left| \lambda_5\right|] >-2~\sqrt{\lambda_1^{'}\lambda_2^{'}}~.
\end{eqnarray}

Assuming that the neutrinos are massless the Yukawa Lagrangian is written as
\begin{equation}\label{LYukawa}
\hspace{-0.3cm}
\begin{array}{rcl}
\mathscr{L}_Y&=&-\dfrac{1}{\sqrt{2}}\bar{D}\{\kappa^D\sin(\beta-\alpha)+\rho^D\cos(\beta-\alpha)\}D~h -\dfrac{\dot{\iota}}{\sqrt{2}}\bar{D}\gamma_5\rho^D D~A \\ && 
-\dfrac{1}{\sqrt{2}}\bar{D}\{\kappa^D\cos(\beta-\alpha)-\rho^D\sin(\beta-\alpha)\}D~H+\dfrac{\dot{\iota}}{\sqrt{2}}\bar{U}\gamma_5\rho^U U~A \\ && -\dfrac{1}{\sqrt{2}}\bar{U}\{\kappa^U\sin(\beta-\alpha)+\rho^U\cos(\beta-\alpha)\}U~h 
-\dfrac{1}{\sqrt{2}}\bar{U}\left\{\kappa^U\cos(\beta-\alpha)\right. \\ && -\left.\rho^U\sin(\beta-\alpha)\right\}U~H  
-\dfrac{1}{\sqrt{2}}\bar{L}\{\kappa^L\sin(\beta-\alpha)+\rho^L\cos(\beta-\alpha)\}L~h \\ && -\dfrac{\dot{\iota}}{\sqrt{2}}\bar{L}\gamma_5\rho^L L~A 
-\dfrac{1}{\sqrt{2}}\bar{L}\{\kappa^L\cos(\beta-\alpha)-\rho^L\sin(\beta-\alpha)\}L~H \\ && 
-\Big[\bar{U}\left(V_{CKM}\rho^DP_R-\rho^UV_{CKM}P_L\right)D~H^{+} 
+\bar{\nu}\rho^LP_RL~H^{+}+h.c.\Big]. \\ && 
\end{array}
\end{equation}
where $\kappa$ is the $3\times3$ diagonal matrix related to the masses by the relation $\kappa^F=\sqrt{2}M^F/\nu$, $M^F$ being the mass matrices of the fermions $D,~U,~L;$ and $P_{R/L}=(\pm\gamma_5)/2$ are the usual parity operators. The different types of Yukawa couplings are selected by choosing the arbitrarily $\rho^F$ accordingly. The CP conservation implies that $\rho^F$ must be symmetric. The off-diagonal elements induce the Higgs boson mediated flavour changing neutral currents. Thus they are strongly constrained. Below we give the table to summarize the values of $\rho^F$ for different types of Yukawa coupling.
\begin{table}[ph]
	\tbl{Different types of Yukawa interaction for 2HDM}
	{\begin{tabular}{@{}ccccc@{}} \toprule
			 & Type I & Type II & Type III & Type IV \\
			\colrule
			$\rho_D$ & $\quad\kappa^D\cot\beta$ & $-\kappa^D\tan\beta$ & $-\kappa^D\tan\beta$ & $\quad\kappa^D\cot\beta$ \\ 
			$\rho_U$ & $\quad\kappa^U\cot\beta$ & $\quad\kappa^U\cot\beta$ & $\quad\kappa^U\cot\beta$ & $\quad\kappa^U\cot\beta$ \\ 
			$\rho_L$ & $\quad\kappa^L\cot\beta$ & $-\kappa^L\tan\beta$ & $\quad\kappa^L\cot\beta$ & $-\kappa^L\tan\beta$ \\ \botrule
		\end{tabular} \label{ta1}}
\end{table}

Since in the presented dark energy model, Higgs fields $h$, $A$ and $H^{\pm}$ are the dark energy fields they must couple to the fermions very weakly or not at all. Just looking at the above table of Yukawa interactions we see that this condition could only be achieved in \emph{only Type I Yukawa interactions} by imposing $\rho^{D/U/L}\ll 1$ and $\tan(\beta-\alpha)=-\cot \beta$.
\subsection{Minimizing the Higgs potential}\label{MtHp}
The extrema of the potential are found by taking
\begin{align}\begin{split}\label{extremaconditions}
&\dfrac{\partial V_H}{\partial \phi_1} {\bigg|}_{\substack{
		\phi_1=\left\langle \phi_1\right\rangle \\
		\phi_2=\left\langle \phi_2\right\rangle
	}}=\dfrac{\partial V_H}{\partial \phi_1^\dagger}\bigg |_{\substack{
	\phi_1=\left\langle \phi_1\right\rangle \\
	\phi_2=\left\langle \phi_2\right\rangle
}}=0~, 
\text{\qquad\space} \dfrac{\partial V_H}{\partial \phi_2}\bigg |_{\substack{
\phi_1=\left\langle \phi_1\right\rangle \\
\phi_2=\left\langle \phi_2\right\rangle
}}=\dfrac{\partial V_H}{\partial \phi_2^\dagger}\bigg |_{\substack{
\phi_1=\left\langle \phi_1\right\rangle \\
\phi_2=\left\langle \phi_2\right\rangle
}}=0~.
\end{split}\end{align}
The most general solution of the conditions (\ref{extremaconditions}) is
\begin{equation*}\label{VeV}
\left\langle \phi_1\right\rangle =\frac{1}{\sqrt{2}}\begin{pmatrix}
0 \\ \nu_1
\end{pmatrix}
\text{\qquad and \qquad}
\left\langle \phi_2\right\rangle =\frac{1}{\sqrt{2}}\begin{pmatrix}
u \\ \nu_2
\end{pmatrix}.
\end{equation*}
The first solution of extrema has been taken to be similar to the Higgs vacuum in the SM and the second one is the most general which can occur. One needs to keep in mind that now $\nu^2=\nu_1^2+\left| \nu_2^2\right|+u^2$. 

When $u\neq 0$, the non-zero value of $u$ will contribute to the ``charged'' type dark energy, which has not been observed. To avoid this, we would take $u=0$. From the extremal conditions given by eq. (\ref{extremaconditions}), we can determine the values of $\nu_1$ and $\nu_2$ \cite{2hdmc,Ginzburg-1}, solving eq. (\ref{extremaconditions}) for the potential given by eq. (\ref{VH}) leads to
\begin{align*}\begin{split}
&\nu_1 {\Big[} 2 \rho_1 \Lambda_1 m_{11}^2\exp{\Big(}\dfrac{\Lambda_1 m_{11}^2}{2}\nu_1^2{\Big)}+\rho_3\Lambda_3\lambda_1\nu_1^2\exp{\Big(}\dfrac{\Lambda_3\lambda_1}{8}\nu_1^4{\Big)}
\\ & \text{\qquad\qquad}-2\text{Re}(m_{12}^2)\frac{\nu_2}{\nu_1}+(\lambda_3+\lambda_4+\text{Re}(\lambda_5))\nu_2^2+3\text{Re}(\lambda_6)\nu_1\nu_2+\text{Re}(\lambda_7)\dfrac{\nu_2^3}{\nu_1}{\Big]}=0~,
\end{split}\end{align*}
and
\begin{align*}\begin{split}
& \nu_2{\Big[}2\rho_2 \Lambda_2 m_{22}^2\exp{\Big(}\dfrac{\Lambda_2 m_{22}^2}{2}\nu_2^2{\Big)}+\rho_4\Lambda_4\lambda_2\nu_2^2\exp{\Big(}\dfrac{\Lambda_4\lambda_2}{8}\nu_2^4{\Big)}  \\ & \text{\qquad\qquad}-2\text{Re}(m_{12}^2)\frac{\nu_1}{\nu_2}+ (\lambda_3+\lambda_4+\text{Re}(\lambda_5))\nu_1^2+\text{Re}(\lambda_6)\dfrac{\nu_1^3}{\nu_2}+3\text{Re}(\lambda_7)\nu_1\nu_2{\Big]}=0~.
\end{split}\end{align*}

The solution of the above equations for $\nu_1$ and $\nu_2$ is practically impossible. 
The  VeV's are approximated by truncating the potential eq. (\ref{VH}) upto forth order \cite{Ginzburg-2}. Applying minimization conditions given by eq. (\ref{extremaconditions}) we get,
\begin{align}
\begin{split}\label{1stcondition}
\nu_1\Bigg[ 2 \rho_1 \Lambda_1 m_{11}^2 & +2\lambda_1^{'}\nu_1^2-2\text{Re}(m_{12}^2)\dfrac{\nu_2}{\nu_1}+ (\lambda_3+\lambda_4+\text{Re}(\lambda_5))\nu_2^2 \\ & \text{\qquad\qquad\qquad\qquad\qquad\space\space} + 3\text{Re}(\lambda_6)\nu_1\nu_2+\text{Re}(\lambda_7)\dfrac{\nu_2^3}{\nu_1}\Bigg]=0,
\end{split}
\end{align}
\begin{align}
\begin{split}\label{2ndcondition}
\nu_2\Bigg[ 2 \rho_2 \Lambda_2 m_{22}^2 & +2\lambda_2^{'}\nu_2^2-2\text{Re}(m_{12}^2)\dfrac{\nu_1}{\nu_2} +(\lambda_3+\lambda_4+\text{Re}(\lambda_5))\nu_1^2 \\ &
\text{\qquad\qquad\qquad\qquad\qquad\space\space}+ \text{Re}(\lambda_6)\dfrac{\nu_1^3}{\nu_2}+3\text{Re}(\lambda_7)\nu_1\nu_2\Bigg]=0~.
\end{split}
\end{align}


With exact $\mathbb{Z}_2$ symmetry the lightest Higgs field is stable. Thus it will be a dark energy candidate. Thus we have \cite{Ginzburg-2},
\begin{equation}\label{Z2Symmetry}
\chi=m_{12}^2=\lambda_6=\lambda_7=0~,
\end{equation}
Using eqs. (\ref{1stcondition}, \ref{2ndcondition}, \ref{Z2Symmetry}), we get four solutions for $\nu_1$ and $\nu_2$, which are
\begin{eqnarray}
\nu_1^2 &= 0 ~, \text{\qquad\qquad\qquad\qquad\qquad} \nu_2^2 &= 0 ~, \label{electroweaksymmetricvacuum}
\\
\nu_1^2 &= 0 ~, \text{\qquad\qquad\qquad\qquad\qquad} \nu_2^2 &= -\dfrac{\rho_2 \Lambda_2 m_{22}^2}{\lambda_2^{'}} ~, \label{inertvacuum1}
\\
\nu_1^2 &= -\dfrac{\rho_1 \Lambda_1 m_{11}^2}{\lambda_1^{'}} ~, \text{\qquad\qquad\qquad} \nu_2^2 &= 0 ~, \label{inertvacuum2}
\end{eqnarray}
\begin{align}
\begin{split}\label{mixedvacuum}
\nu_1^2 &= -\dfrac{2(2\rho_1 \Lambda_1\lambda_2^{'} m_{11}^2-\rho_2 \Lambda_2 \lambda_{345}m_{22}^2)}{4\lambda_1^{'} \lambda_2^{'}-\lambda_{345}^2} ~, 
\text{\quad} \nu_2^2 = -\dfrac{2(2\rho_2 \Lambda_2 \lambda_1^{'} m_{22}^2-\rho_1 \Lambda_1\lambda_{345}m_{11}^2)}{4\lambda_1^{'} \lambda_2^{'}-\lambda_{345}^2} ~,
\end{split}
\end{align}
where $\lambda_{345}=\lambda_3+\lambda_4+\text{Re}(\lambda_5)$.

After both symmetries are broken, the Lagrangian violates the $\mathbb{Z}_2$ symmetry given by eqs. (\ref{1stZ2symmetry}) and (\ref{2ndZ2symmetry}).
The physical fields in this model are a combination of fields from $\phi_1$ and $\phi_2$. The complete theory and phenomenology of this model is discussed in \cite{Brancoetal}. This type of 
solution is given by eq. (\ref{mixedvacuum}) and the masses of the Higgs fields in this vacuum state are given by eq. (\ref{mA}), (\ref{mh+}) and (\ref{mHh}).
\section{Phase transitions in the two Higgs doublet model}
When we look for dark energy to be some physical field(s), then it becomes essential to also look for its evolution in the history (cooling) of the Universe.

In the quantum field theory at non-zero temperature the terms $m_{ii}^2$ of the quadratic terms evolve with temperature as \cite{Ginzburg-1}
\begin{eqnarray}
&& m_{11}^2\longrightarrow m_{11}^2+\dfrac{1}{2}c_1 T^2, \label{m112thermalevolution} \\ &&
m_{22}^2\longrightarrow m_{22}^2+\dfrac{1}{2}c_2 T^2, \label{m222thermalevolution}
\end{eqnarray}
where
$$c_1=\dfrac{6\lambda_1^{'}+2\lambda_3+\lambda_4}{12}+\dfrac{3g_1^2+g_1'^2}{32}+\dfrac{g_t^2+g_b^2}{8},$$
$$c_2=\dfrac{6\lambda_2^{'}+2\lambda_3+\lambda_4}{12}+\dfrac{3g_2^2+g_2'^2}{32},$$
and $g_{1,2}$ and $g'_{1,2}$ are the Electroweak gauge couplings with doublet $\phi_1$ and $\phi_2$ respectively, $g_t\approx 1$ and $g_b\approx 0.03$ are the top and bottom quark Yukawa couplings with the doublet $\phi_1$ in inert doublet model respectively.
In general, $c_1$ and $c_2$ can have any sign but the potential positivity implies that (in any situation)
\begin{equation*}
c_1+c_2>0,
\end{equation*}
with the above mentioned positivity constraints. The case when $c_1,c_2>0$ will now be considered.

The Higgs potential with the new quadratic terms now becomes
\begin{align}
\begin{split}\label{VHT}
&\hspace{-0.4cm}V_H(\phi_1,\phi_2,T)=E^{'}_{vac}+\dfrac{1}{2}c_1(T^2-{T_c}_1^2)(\phi_1^\dagger\phi_1)+\dfrac{1}{2}c_2(T^2-{T_c}_2^2)(\phi_2^\dagger\phi_2) 
\\ &\hspace{-0.4cm} 
+{m_{12}^2}(\phi_1^\dagger\phi_2)+{m_{12}^2}^{*}(\phi_2^\dagger\phi_1)
+\lambda_1^{'}(\phi_1^\dagger\phi_1)^2+\lambda_2^{'}(\phi_2^\dagger\phi_2)^2+\lambda_3 (\phi_1^\dagger\phi_1)(\phi_2^\dagger\phi_2) 
\\ &\hspace{-0.4cm}
+\lambda_4(\phi_1^\dagger\phi_2)(\phi_2^\dagger\phi_1)+\dfrac{1}{2}[\lambda_5(\phi_1^\dagger\phi_2)^2 +\lambda_5^* (\phi_2^\dagger\phi_1)^2] 
+(\phi_1^\dagger\phi_1)[ \lambda_6(\phi_1^\dagger\phi_2)+\lambda_6^*(\phi_2^\dagger\phi_1) ] 
\\ &\hspace{-0.4cm}
+(\phi_2^\dagger\phi_2)[ \lambda_7(\phi_1^\dagger\phi_2)+\lambda_7^*(\phi_2^\dagger\phi_1) ]+\textit{higher order terms}.
\end{split}
\end{align}
In this case corresponding to the two different VeV's of the fields $\phi_1$ and $\phi_2$, there would be two critical temperatures, $T_{c_{1}}=\sqrt{-\dfrac{2~\rho_1 \Lambda_1 m_{11}^2}{c_1}}$ and $T_{c_{2}}=\sqrt{-\dfrac{2~\rho_2 \Lambda_2 m_{22}^2}{c_2}}$ respectively. In this analysis, we require that the phase transitions in doublet $\phi_2$ occur after phase transitions in $\phi_1$, thus $T_{c_{1}}>T_{c_{2}}$.
When $T<T_{c_{2}}$ symmetry broke spontaneously in both $\phi_1$ and $\phi_2$ at ${T_c}_1$ and ${T_c}_2$ respectively. Now, both fields are perturbed around their true minima. This perturbation give masses to all the physical mixed fields of the doublets $\phi_1$ and $\phi_2$.
\subsection*{Constraint on parameters from phase transitions}
Here we require that the phase transitions in the doublet $\phi_1$ (the new SM Higgs doublet) occur at the same temperature as the usual SM Higgs doublet. The critical temperature at which phase transitions occur in the SM is \cite{kapusta},
$$T_c^2=\dfrac{4~\lambda~\nu^2}{2\lambda+\dfrac{3}{4}g^2+\dfrac{1}{4}{g^{'}}^2},$$
\\
where $\lambda=0.1305$ is the quartic coupling of the Higgs doublet in the SM, $g=0.6376$ is the Higgs doublet coupling with the SU(2) gauge group and $g^{'}=0.3441$ is the Higgs doublet coupling with U(1) gauge group. With the values of parameters given above we get $T_c=230.3186$GeV.
Setting $T_{c_1}=T_c$ gives 
\begin{equation}\label{phasetransitionsconstraint}
2\lambda_3+\lambda_4=0.7869~.
\end{equation}
In deriving the above equation, $\lambda_1^{'}=0.1305$, $g_1=0.6376$, $g_1'=0.3441$, $g_t=1$ and $g_b=0.03$ have been used.
\section{Higgs field(s) as dark energy}
The Universe is homogeneous and isotropic at the cosmological scale, and its dynamics is described by the Friedmann equations given by eqs. (\ref{1stFriedmannequation}) and (\ref{2ndFriemannequation}). Equation (\ref{2ndFriemannequation}) says that accelerated expansion will occur when $\omega_{\text{eff}}<-\frac{1}{3}$. For the field $\phi_2$ to be the dark energy field, it must give $\omega_{\text{eff}}<-\frac{1}{3}$ at $Z\approx 0.37$ (with $\Omega_\text{m}=0.3$ and $\Omega_{\text{DE}}=0.7$). For this purpose, we need to solve the Euler-Lagrange equations, which are
\begin{eqnarray}\label{EulerLagrangeEquations}
\partial_\mu\left( \dfrac{\partial(\sqrt{-g}{\mathscr{L}_{Higgs})}}{\partial(\partial_\mu \psi_i)} \right)-\dfrac{\partial(\sqrt{-g}{\mathscr{L}_{Higgs}})}{\partial \psi_i}=0,
\end{eqnarray}
where $\psi_i$ are different fields of doublets $\phi_1$ and $\phi_2$.

The Euler-Lagrange equations of motion in the FRW Universe $(\sqrt{-g}=a(t)^3)$ for the fields $H^C$, $h$ and $A$ are given in \ref{app:ELeq}. The energy density and pressure after expansion of 2HDM Higgs Lagrangian for physical fields are given in \ref{app:rhoPHiggs}. For the cosmological evolution of the fields $h$, $A$ and $H^C$, the Euler-Lagrange equations are solved with the Friedmann equations numerically in the flat Universe ($\kappa=0$). The initial conditions used are $h_{ini}=A_{ini}=M_P$ and  ${H^C}_{ini}={\dot{h}}_{ini}={\dot{A}}_{ini}={\dot{H}^C}_{ini}=0$. The Higgs fields masses in this analysis were calculated using eq. (\ref{mA}), (\ref{mh+}) and (\ref{mHh}) with $m_{H}=m_{H_{SM}}=125.7$GeV. Note that after imposing the $\mathbb{Z}_2$ symmetry, $m_{H}=m_{H_{SM}}$, vacuum energy density constraints and constraint from phase transitions eq. (\ref{phasetransitionsconstraint}) there are seven parameters which determine the masses of Higgs fields. 
We then made some arbitrary choice of parameters \footnote{The charged Higgs mass was chosen $\gtrsim 80$GeV as suggested by Particle date group (PDG) \cite{PDG}.} since there were more parameters than equations to determine all unknowns.
The masses of Higgs bosons in the analysis are taken to be
\begin{equation*}
m_{h}=10^{-60}\text{GeV,\quad} m_{A}=0\text{\space GeV,\quad} m_{H^C}=154.305\text{GeV}.
\end{equation*}

The Universe in our model evolves in the mixed vacuum state. In this situation, phase transition in the second doublet has occurred, so we can have either an ever accelerating expansion or accelerated expansion for a limited time depending upon the Yukawa couplings. We are especially interested in the case of an ever accelerating expansion.

After solving eqs. (\ref{hequationofmotion}, \ref{Aequationofmotion}, \ref{chargedHiggsequationofmotion}) along with the Friedmann equations, we find that
the Higgs fields in our model do not evolve as the Universe get older 
%
%
and, $\omega_{Higgs}$ does not deviate from $-1$. Since, $\omega_{Higgs}<-1/3$ all the time in evolution after matter, radiation equality once the Higgs energy density starts to dominate, we observe an accelerated expansion of the Universe.

\begin{figure*}[h!]
	\begin{center}
		\centering
		\includegraphics[width=\textwidth]{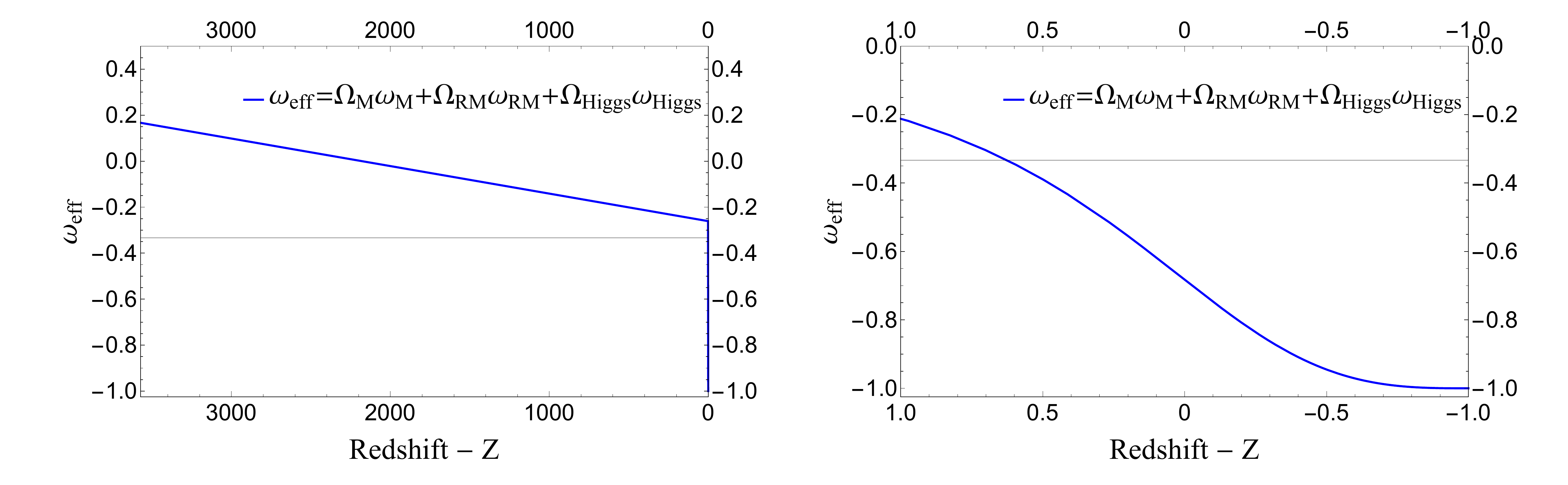}
		\caption{Effective equation of state $\omega_{eff}=\Omega_{DE}\omega_{DE}+\Omega_{R}\omega_R+\Omega_{M}\omega_M$.}
		\label{fig:Omegaeffective}
	\end{center}
\end{figure*}

\begin{figure*}[h!]
	\centering
	\includegraphics[scale=0.52]{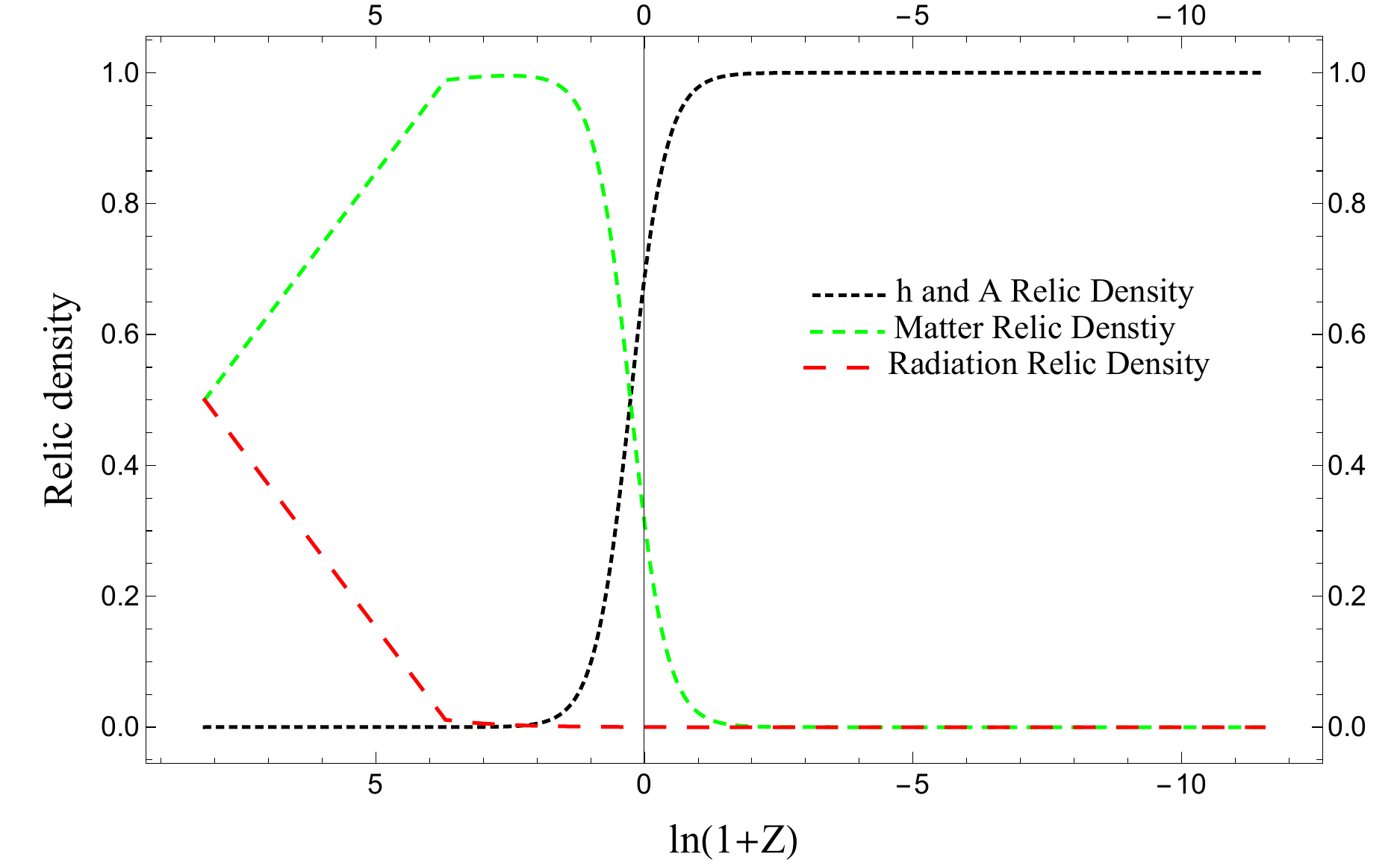}
	\caption{Relic densities of different components as a function of $ln$[$a_0/a$].}
	\label{fig:relicdensities1}
\end{figure*}

\begin{figure*}[h!]
	\centering
	\includegraphics[scale=0.52]{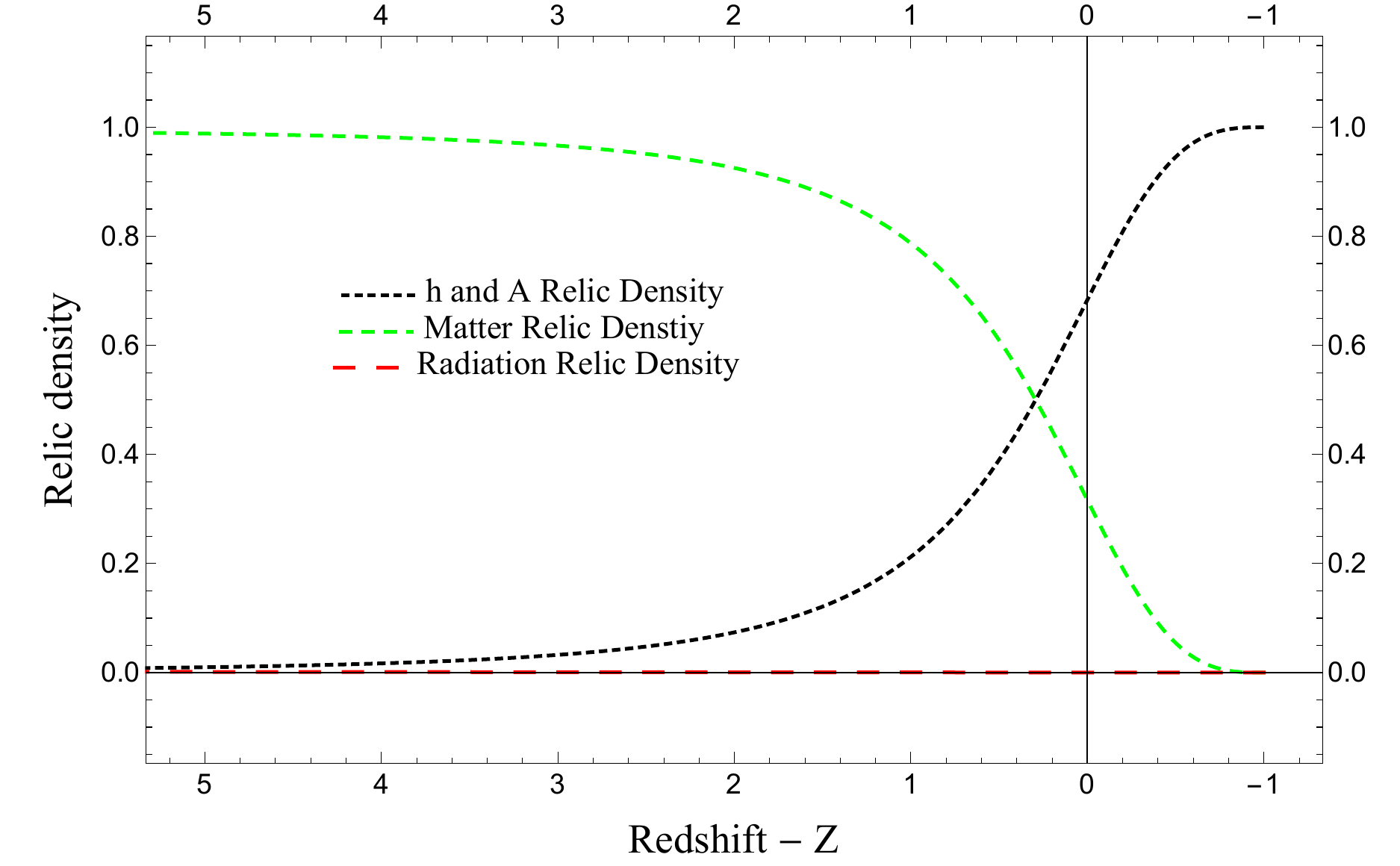}
	\caption{Relic densities of different components as a function of redshift.}
	\label{fig:relicdensities2}
\end{figure*}

The $\omega_{\text{eff}}$ in fig. (\ref{fig:Omegaeffective}) starts from $\approx0.167$ (set by initial conditions $\Omega_{{Higgs}_{int}}=0$ and $\Omega_{{NR}_{int}}=\Omega_{R_{int}}=0.5$
) and decreases as $\Omega_R$ decreases. The effective EoS parameter $\omega_{\text{eff}}$ comes down to $-1/3$ at $Z=0.63$ 
in our numerical solution shown in fig. (\ref{fig:Omegaeffective}). Prior to this time, non-relativistic matter dominates and the Universe decelerates at the highest rate ($\Omega_{NR}\approx1$) as attraction dominates the Higgs repulsion. After that $\omega_{\text{eff}}$ starts to decrease as $\Omega_{NR}$ decreases and the Higgs relic energy density increases as shown in fig. (\ref{fig:relicdensities1}) and (\ref{fig:relicdensities2}). From this time on the Higgs negative pressure dominates forever and $\omega_{\text{eff}}$ eventually settles down to $-1$.	

In Fig. (\ref{fig:relicdensities1}) $\Omega_R$, $\Omega_{NR}$  and $\Omega_H$ are plotted from $t_{int}$ asymptotically to infinity with $\Omega_H$ only now beginning to dominate. Figure (\ref{fig:relicdensities2}) is plotted for $\Omega_R$, $\Omega_{NR}$  and $\Omega_H$ against the redshift $Z$ from $Z\approxeq 5$ to $Z=-1$. Around $Z\approx 5$ the dark energy Higgs has started to become prominent. 
For $Z < 0.63$ 
dark energy Higgs dominates causing the accelerated expansion of the Universe.
\section{Conclusion}\label{Conclusion}

Here we demonstrated that it is not necessary to go outside
the standard model of Particle Physics to provide a dark energy candidate.
Since the Higgs sector is actually unconstrained and it is restricted to the
minimal Higgs sector only for definiteness, one could introduce a second Higgs that is not inconsistent with observation. In particular, we take a two
Higgs doublet model for the purpose. We find that if the present Universe is
described by the vacuum given by 2HDM then the component scalar fields of the
2HDM can be one possible candidate for the dark energy. Since the present
contribution of the dark energy to the critical energy density is about
$0.7$, this value is obtained by taking the masses of CP-even field(s) very
small. The most important thing is that with the initial conditions, set the
mass of the charged field ($\phi_2^c$) is unconstrained 
if we remove the parametric constraint given by eq.  (\ref{phasetransitionsconstraint}), which
is obtained using the phase transitions bound, this model will fit for any value
of mass of $\phi_2^c$. The initial conditions for the charged field were taken in accordance with the observation that the dark energy (vacuum) is not charged.

The $\mathbb{Z}_2$ symmetric potential in our model ensures that the lightest Higgs, $\eta_2$ and $\chi_2$, do not decay into any other Higgs. 
On the other hand, the coupling of the dark energy Higgs to the fermions can only be suppressed by choosing 
\emph{Type I Yukawa interactions} with $\rho^{D/U/L}\ll 1$ and $\tan(\beta-\alpha)=-\cot \beta$.

To suppress the interaction of the Higgs fields $\eta_2$, $\chi_2$ and $\phi_2^\pm$ with the gauge bosons $W^{\pm}$, $Z$ and $A(\mu)$,  we imposed the condition that the SU(2) doublet $\phi_2$ is very weakly (and differently than $\phi_1$) coupled with the gauge bosons. Thus, here $g_2\ll g_1\text{ and }g_2'\ll g_1'$. This allowed us to conclude that the decay modes that include the Higgs fields $\eta_2$, $\chi_2$ and $\phi_2^\pm$ are negligible as compared to the modes that include the SM Higgs. The couplings $g_1 \text{ and }g_1'$ have the values of the SM Higgs gauge couplings in the SM but the values of $g_2 \text{ and }g_2'$ must be set using the results from LHC or any other Particle Physics experiment.

One thing that remains important to check in all extensions of the SM is whether the Higgs potential contains the vacuum instability or not. If it does, then does it make the vacuum instability worse as compared to the SM. The answer to the question for our model is that although it contains the vacuum instability, due to the coupling of the second Higgs with the SM Higgs which is $\mathcal{O}(10^{-126})$, it will not affect the RGEs running of the SM Higgs. Thus it does not make the vacuum instability worse. We expect the vacuum instability to occur at approximately the same scale as it occurs in the SM.

In conclusion, since we get $\omega_{eff}<-1/3$ after solving the Euler-Lagrange equations numerically, the proposed Higgs field is a possible candidate for the current observed accelerated expansion.
\section*{Acknowledgments}
This work is supported by {\textit{National University of Sciences and Technology (NUST), Sector H-12 Islamabad 44000, Pakistan}} and \emph{Higher Education Commission (HEC) of Pakistan} under the project no. NRPU-3053. AQ is grateful to NUST for support to attend the Workshop and to Prof. Francesco De Paolis for support under INFN to visit the Selento University at Lecce and to Prof. Remo Ruffini for support at ICRANet, Pescara, where this work was largely finalized.
\clearpage
\appendix
\section{Euler-Lagrange equations}\label{app:ELeq}
The Euler Lagrange equation of motion in FRW Universe $(\sqrt{-g}=a(t)^3)$ for $h$ is
{\scriptsize
	\begin{equation}\label{hequationofmotion}
	\left.\hspace{-0.35cm}
	\begin{aligned}
	& h''+3 \dfrac{a'}{a} h'-\Bigg\{-\lambda_{345}\cos^2\alpha\sin^2\alpha\text{ }h^3-\dfrac{1}{2}\nu^3\lambda_{345}\cos\beta\sin\beta\big(\cos\alpha\cos\beta-\sin\alpha\sin\beta\big) 
	\\ &  
	-\dfrac{3}{2}\nu\lambda_{345}\cos\alpha\sin\alpha\big(\sin\alpha\sin\beta-\cos\alpha\cos\beta\big)h^2-\dfrac{1}{2}\nu^2\lambda_{345}\big(\cos^2\alpha\cos^2\beta+\sin^2\alpha\sin^2\beta
	\\ &
	-4\cos\alpha\cos\beta\sin\alpha\sin\beta\big)h-\dfrac{1}{2}\nu\bigg(\lambda_{34-5}\big(\cos\alpha \sin^3\beta-\cos^3\beta\sin\alpha\big)+2\lambda_5\cos\beta\sin\beta
	\\ &
	\big(\sin\alpha\sin\beta-\cos\alpha\cos\beta\big)\bigg)A^2-\nu\bigg(\lambda_3\big(\cos\alpha\sin^3\beta-\cos^3\beta\sin\alpha\big)+\big(\lambda_4+\lambda_5\big)\cos\beta\sin\beta
	\\ &
	\big(\sin\alpha\sin\beta-\cos\alpha\cos\beta\big)\bigg){H^C}^2-2\bigg(A^2\Big(\dfrac{1}{4}\lambda_{34-5}\big(\cos^2\beta\sin^2\alpha+\cos^2\alpha\sin^2\beta\big)
	\\ &
	+\lambda_5\cos\alpha\cos\beta\sin\alpha\sin\beta\Big)+{H^C}^2\Big(\dfrac{1}{2}\lambda_3\big(\cos^2\beta\sin^2\alpha+\cos^2\alpha\sin^2\beta\big)+\big(\lambda_4+\lambda_5\big)
	\\ &
	\cos\alpha\cos\beta\sin\alpha\sin\beta\Big)\bigg)h-\Lambda_1\rho_1\Bigg(\dfrac{1}{2}\nu^3\big(\lambda_1\cos^3\beta\sin\alpha+\lambda_{345}\cos\beta\sin\alpha\sin^2\beta\big)-\dfrac{1}{2}\nu^2
	\\ &
	\big(\lambda_1\cos^2\beta\sin^2\alpha+\lambda_{345}\sin^2\alpha\sin^2\beta\big)h\Bigg)\text{Exp}\Bigg[\Lambda_1\Big(\dfrac{1}{2}\nu^3\big(\cos^3\beta\sin\alpha\lambda_1 
	\\ &
	+\lambda_{345}\cos\beta\sin\alpha\sin^2\beta\big)h-\dfrac{1}{4}\nu^2\big(\cos^2\beta\sin^2\alpha\lambda_1+\lambda_{345}\sin^2\alpha\sin^2\beta\big)h^2\Big)\Bigg]
	\\ &
	\Lambda_2\rho_2\Bigg(-\dfrac{1}{2}\nu^2\big(\lambda_2\cos^2\alpha\sin^2\beta+\lambda_{345}\cos^2\alpha\cos^2\beta\big)h-\dfrac{1}{2}\nu^3\big(\lambda_2\cos\alpha\sin^3\beta
	\\ &
	+\lambda_{345}\cos\alpha\cos^2\beta\sin\beta\big)\Bigg)-\text{Exp}\Bigg[\Lambda_2\Big(-\dfrac{1}{4} \nu^2\big(\lambda_2\cos^2\alpha\sin^2\beta+\lambda_{345}\cos^2\alpha\cos^2\beta\big)h^2
	\\ &
	-\dfrac{1}{2}\nu^3\big(\lambda_2\cos\alpha\sin^3\beta+\lambda_{345}\cos\alpha\cos^2\beta\sin\beta\big)h\Big)\Bigg]-\dfrac{1}{8}\lambda_1\Lambda_3\rho_3\Bigg(-4\nu^3\cos^3\beta\sin\alpha
	\\ &
	+12\nu^2\cos^2\beta\sin^2\alpha h-12\nu\cos\beta\sin^3\alpha h^2+4\sin^4\alpha h^3\Bigg) \text{Exp}\Bigg[\dfrac{1}{8}\lambda_1\Lambda_3\Big(-4\nu^3\cos^3\beta\sin\alpha h
	\\ &
	+6\nu^2\cos^2\beta\sin^2\alpha h^2-4\nu\cos\beta\sin^3\alpha h^3+\sin^4\alpha h^4\Big) \Bigg]-\dfrac{1}{4}\lambda_1\Lambda_3\rho_3\Big(A^2+2{H^C}^2\Big)
	\\ &
	\Big(-2\nu\cos\beta\sin\alpha\sin^2\beta+2\sin^2\alpha\sin^2\beta h\Big) \text{Exp}\Bigg[\dfrac{1}{4}\big(A^2+2{H^C}^2\big)\big(-2\nu\cos\beta h\sin\alpha\sin^2\beta
	\\ &
	+h^2\sin^2\alpha\sin^2\beta\big)\lambda_1\Lambda_3\Bigg]-\dfrac{1}{4}\lambda_2\Lambda_4\rho_4\Big(A^2+2{H^C}^2\Big)\Big(2\cos^2\alpha\cos^2\beta h
	\\ &
	+2\nu\cos\alpha\cos^2\beta\sin\beta\Big)\text{Exp}\Bigg[\dfrac{1}{4}\lambda_2\Lambda_4\big(A^2+2{H^C}^2\big)\big(\cos^2\alpha\cos^2\beta h^2
	\\ & 
	+2\nu\cos\alpha\cos^2\beta h\sin\beta\big)\Bigg]-\dfrac{1}{8}\lambda_2\Lambda_4\rho_4 \Bigg(4\cos^4\alpha h^3+12\nu\cos^3\alpha h^2\sin\beta
	\\ &
	+12\nu^2\cos^2\alpha h\sin^2\beta+4\nu^3\cos\alpha\sin^3\beta\Bigg)\text{Exp}\Bigg[\dfrac{1}{8} \Big(\cos^4\alpha h^4+4\nu\cos^3\alpha h^3\sin\beta
	\\ &
	+6\nu^2\cos^2\alpha h^2\sin^2\beta+4\nu^3\cos\alpha h\sin^3\beta\Big)\lambda_2\Lambda_4\Bigg]
	\Bigg\}=0 
	\end{aligned}
	\right\}
	\qquad \text{}
	\end{equation}
}
\clearpage
\noindent The Euler Lagrange equation of motion in FRW Universe $(\sqrt{-g}=a(t)^3)$ for $A$ is
{\scriptsize
	\begin{equation}\label{Aequationofmotion}
	\left.\hspace{-0.35cm}
	\begin{aligned}
	& A''+3\dfrac{a'}{a}A'-\Bigg\{-\lambda_{345}\cos^2\beta\sin^2\beta A^3-2\lambda_{345}\cos^2\beta\sin^2\beta A{H^C}^2-2 A h^2\bigg(\dfrac{1}{4}\lambda_{34-5}
	\\ &  
	\big(\cos^2\beta  \sin^2\alpha +\cos^2\alpha  \sin^2\beta \big) +\lambda_5\cos\alpha \cos\beta \sin\alpha \sin\beta \bigg) -\dfrac{1}{2} \nu ^2 A \bigg(\lambda_{34-5}\cos^4\beta
	\\ & 
	+\lambda_{34-5}\sin^4\beta-4\lambda_5\cos^2\beta \sin^2\beta \bigg) -\nu A h \bigg(\lambda _{34-5}\big(\cos\alpha \sin^3\beta-\cos^3\beta  \sin\alpha \big)
	\\ &
	+2\lambda_5 \cos\beta \sin\beta \big(\sin\alpha \sin\beta-\cos\alpha \cos\beta\big) \bigg) +\dfrac{1}{2}\nu^2\Lambda_1\rho_1 \sin^2\beta A \left(\cos^2\beta \lambda_1+\lambda_{345}\sin^2\beta \right)
	\\ &
	\text{Exp}\Bigg[-\dfrac{1}{4} \nu ^2 \Lambda _1\big(A^2+2 {H^C}^2\big) \sin^2\beta \big(\lambda_1\cos^2\beta+\lambda_{345}\sin^2\beta \big) \Bigg]+\dfrac{1}{2}\nu ^2\Lambda _2 \rho _2 \cos^2\beta A 
	\\ & 
	\Big(\lambda _2\sin^2\beta +\lambda_{345}\cos^2\beta \Big)  \text{Exp}\Bigg[-\dfrac{1}{4} \nu ^2\Lambda _2 \cos^2\beta  \bigg(A^2+2 {H^C}^2\bigg) \bigg(\lambda _2\sin^2\beta +\lambda _{345}\cos^2\beta  \bigg) \Bigg] 
	\\ & 
	-\dfrac{1}{2}\lambda_1 \Lambda _3 \rho _3 A \bigg(-2 \nu  \cos\beta \sin\alpha \sin^2\beta h+ \sin^2\alpha\sin^2\beta h^2\bigg) \text{Exp}\Bigg[\dfrac{1}{4}\lambda_1 \Lambda _3 \Big(A^2+2 {H^C}^2\Big)
	\\ &
	\Big(-2 \nu \cos\beta \sin\alpha \sin^2\beta h+ \sin^2\alpha  \sin^2\beta h^2\Big) \Bigg] -\dfrac{1}{8}\lambda_1 \Lambda _3 \rho _3 \bigg(4 \nu ^2 A \cos^2\beta  \sin^2\beta +4 A^3 \sin^4\beta
	\\ & 
	+8 A {H^C}^2 \sin^4\beta\bigg)\text{Exp}\Bigg[\dfrac{1}{8}\lambda_1 \Lambda _3 \bigg(2 \nu ^2 \cos^2\beta  \Big(A^2+2 {H^C}^2\Big) \sin^2\beta +4 A^2 {H^C}^2 \sin^4\beta 
	\\ &
	+\Big(A^4+4{H^C}^4\Big)\sin^4\beta \bigg)\Bigg]-\dfrac{1}{2}\lambda _2 \Lambda _4 \rho _4 A \bigg(\cos^2\alpha  \cos^2\beta  h^2+2 \nu  \cos\alpha \cos^2\beta \sin\beta h\bigg)
	\\ &
	\text{Exp}\Bigg[\dfrac{1}{4} \Big(A^2+2 {H^C}^2\Big) \Big(\cos^2\alpha  \cos^2\beta  h^2+2 \nu  \cos\alpha \cos^2\beta h \sin\beta\Big) \lambda _2 \Lambda _4\Bigg]   
	\\ & 
	-\dfrac{1}{8}\lambda _2 \Lambda _4 \rho _4 \bigg(4 A^3 \cos^4\beta +8 A \cos^4\beta  {H^C}^2+4 \nu ^2 A \cos^2\beta\sin^2\beta \bigg) \text{Exp}\Bigg[\dfrac{1}{8} \bigg(4 A^2 \cos^4\beta  {H^C}^2
	\\ &
	+\cos^4\beta  \Big(A^4+4 {H^C}^4\Big)+2 \nu ^2 \cos^2\beta
	\Big(A^2+2 {H^C}^2\Big)\sin^2\beta \bigg) \lambda _2 \Lambda _4\Bigg] \Bigg\}=0
	\end{aligned}
	\right\}
	\qquad \text{}
	\end{equation}
}
\clearpage
\noindent The Euler Lagrange equation of motion in FRW Universe $(\sqrt{-g}=a(t)^3)$ for $H^C$ is
{\scriptsize
	\begin{equation}\label{chargedHiggsequationofmotion}
	\left.\hspace{-0.35cm}
	\begin{aligned}
	& 2 \big(H^C\big)''+6 \dfrac{a'}{a}\big(H^C\big)'-\Bigg\{-2\lambda_{345} A^2 \cos^2\beta  H^C \sin^2\beta -4\lambda_{345} \cos^2\beta \sin^2\beta {H^C}^3 
	\\ & 
	-2 h^2 H^C \bigg(\dfrac{1}{2} \lambda _3\big(\cos^2\beta  \sin^2\alpha +\cos^2\alpha  \sin^2\beta \big) +\big(\lambda_4+\lambda_5\big)\cos\alpha\cos\beta\sin\alpha\sin\beta\bigg)
	\\ &
	-\nu^2H^C\bigg(\lambda_3\big(\cos^4\beta+\sin^4\beta\big)-2\big(\lambda_4+\lambda_5\big)\cos^2\beta\sin^2\beta\bigg) -2\nu  h H^C \Bigg(\lambda_3\big(-\cos^3\beta  \sin\alpha 
	\\ & 
	+\cos\alpha \sin^3\beta \big) +\big(\lambda _4+\lambda_5\big)
	\cos\beta \sin\beta \big(-\cos\alpha \cos\beta+\sin\alpha \sin\beta\big)\Bigg)+\nu ^2 \Lambda _1 \rho _1 H^C \sin^2\beta
	\\ &
	\Big(\lambda_1\cos^2\beta+\lambda_{345}\sin^2\beta\Big)\text{Exp}\Bigg[-\dfrac{1}{4}\Lambda_1\nu^2 \sin^2\beta\Big(A^2+2 {H^C}^2\Big)  \Big(\lambda_1\cos^2\beta+\lambda_{345}\sin^2\beta\Big) \Bigg] 
	\\ &
	+\nu ^2 \Lambda _2 \rho _2 \cos^2\beta  H^C \Big(\lambda _2\sin^2\beta +\lambda_{345}\cos^2\beta \Big)\text{Exp}\Bigg[-\dfrac{1}{4}\Lambda_2 \nu ^2 \cos^2\beta  \Big(A^2+2 {H^C}^2\Big)
	\\ &
	\Big(\lambda_2\sin^2\beta +\lambda_{345}\cos^2\beta \Big) \Bigg] -\lambda_1 \Lambda _3 \rho _3 H^C \Big(-2 \nu  \cos\beta \sin\alpha \sin^2\beta h+ \sin^2\alpha \sin^2\beta h^2\Big)
	\\ &
	\text{Exp}\Bigg[\dfrac{1}{4}\lambda_1 \Lambda _3 \Big(A^2+2 {H^C}^2\Big) \Big(-2 \nu \cos\beta \sin\alpha \sin^2\beta h+ \sin^2\alpha\sin^2\beta h^2\Big)\Bigg] 
	\\ &
	-\dfrac{1}{8}\lambda_1 \Lambda _3 \rho _3\Big(8 \nu ^2 \cos^2\beta  H^C \sin^2\beta +8 A^2 H^C \sin^4\beta +16 {H^C}^3 \sin^4\beta \Big)
	\\ &
	\text{Exp}\Bigg[\dfrac{1}{8}\lambda_1 \Lambda _3 \Big(2 \nu ^2 \cos^2\beta \sin^2\beta \big(A^2+2 {H^C}^2\big)+4 A^2 {H^C}^2 \sin^4\beta +\sin^4\beta\big(A^4+4{H^C}^4\big) \Big)\Bigg]
	\\ &
	-\lambda _2 \Lambda _4 \rho _4 H^C \bigg(\cos^2\alpha  \cos^2\beta  h^2+2 \nu  \cos\alpha \cos^2\beta  h \sin\beta\bigg)\text{Exp}\Bigg[\dfrac{1}{4} \Big(A^2+2 {H^C}^2\Big)
	\\ &
	\Big(\cos^2\alpha  \cos^2\beta  h^2+2 \nu  \cos\alpha \cos^2\beta  h \sin\beta\Big) \lambda _2 \Lambda _4\Bigg] -\dfrac{1}{8}\lambda _2 \Lambda _4 \rho _4\bigg(8 A^2 \cos^4\beta  H^C
	\\ &
	+16 \cos^4\beta  {H^C}^3+8 \nu ^2 \cos^2\beta H^C \sin^2\beta \bigg) \text{Exp}\Bigg[\dfrac{1}{8}\lambda _2 \Lambda _4 \bigg(4 A^2 \cos^4\beta  {H^C}^2
	\\ &
	+\cos^4\beta \Big(A^4+4 {H^C}^4\Big)+2 \nu ^2 \cos^2\beta\Big(A^2+2 {H^C}^2\Big) \sin^2\beta \bigg) \Bigg]
	\Bigg\}=0
	\end{aligned}
	\right\}
	\qquad \text{}
	\end{equation}
}
where `C' instead of $+$ or $-$.
\newpage
\section{Energy density and Pressure}\label{app:rhoPHiggs}
The energy density and pressure after expansion of 2HDM Higgs Lagrangian for physical fields is
{\scriptsize
	\begin{equation}\label{rhoHiggsPHiggs}
	\left.\hspace{-0.35cm}
	\begin{aligned}
	&\rho_{Higgs}/P_{Higgs} = \dfrac{1}{4} \nu ^4\lambda_{345} \cos^2\beta  \sin^2\beta +\rho _1\text{Exp}\bigg[-\dfrac{1}{4}\Lambda_1\nu^4\Big(\lambda_1\cos^4\beta 
	\\ &
	+\lambda_{345}\cos^2\beta\sin^2\beta\Big)\bigg] 
	+\rho_2\text{Exp}\bigg[-\dfrac{1}{4}\Lambda_2\nu^4\Big(\lambda_2\sin^4\beta+\lambda_{345}\cos^2\beta\sin^2\beta\Big)\bigg]
	\\ &
	+\rho_3\text{Exp}\bigg[\dfrac{1}{8}\Lambda _3\lambda_1\nu ^4 \cos^4\beta\bigg]+\rho_4\text{Exp}\bigg[\dfrac{1}{8}\Lambda_4\lambda_2\nu^4\sin^4\beta\bigg]+\rho_1\text{Exp}\bigg[-\dfrac{1}{4}\Lambda_1\nu^2\sin^2\beta
	\\ &
	\Big(\lambda_1\cos^2\beta+\lambda_{345}\sin^2\beta\Big)\Big(2{H^C}^2 +A^2\Big)\bigg]+\rho_2\text{Exp}\bigg[-\dfrac{1}{4}\nu^2\Lambda_2\cos^2\beta\Big(2{H^C}^2+A^2\Big)
	\\ &
	\Big(\lambda_2\sin^2\beta+\lambda_{345}\cos^2\beta\Big)\bigg]+\rho _3\text{Exp}\bigg[\dfrac{1}{8}\Lambda _3\lambda_1\Big(\big(A^4+4 {H^C}^4\big)\sin^4\beta+2\nu ^2 \cos^2\beta \sin^2\beta
	\\ &
	\big(A^2+2{H^C}^2\big)+4\sin ^4\beta{H^C}^2A^2\Big)\bigg]
	+\rho _4\text{Exp}\bigg[\dfrac{1}{8}\Lambda _4\lambda _2\Big(\big(A^4+4 {H^C}^4\big)\cos^4\beta
	\\ &
	+2 \nu ^2 \cos^2\beta  \sin^2\beta \big(A^2+2{H^C}^2\big)
	+4\cos ^4\beta{H^C}^2A^2\Big)\Bigg] +\dfrac{1}{4}\lambda_{345} \cos^2\beta  \sin ^2\text{$\beta $A}^4
	\\ & 
	+\lambda_{345}\cos^2\beta\sin^2\beta{H^C}^4+\dfrac{1}{4}\nu^2A^2\bigg(\lambda_{34-5}\cos^4\beta+\lambda_{34-5}\sin^4\beta-4\lambda_5\cos^2\beta\sin^2\beta\bigg) 
	\\ & 
	+\dfrac{1}{2}\nu^2{H^C}^2\bigg(\lambda_3\Big(\cos^4\beta+\sin^4\beta\Big)-2\cos^2\beta\sin^2\beta\Big(\lambda _4+\lambda_5\Big)\bigg)+\lambda_{345}\cos^2\beta\sin^2\beta A^2{H^C}^2
	\\ &
	+\rho _1\text{Exp}\Bigg[\Lambda _1\bigg(-\frac{1}{4}\nu ^2 h^2
	\Big(\cos^2\beta  \sin^2\alpha  \lambda_1+\sin^2\alpha  \sin^2\beta \lambda_{345}\Big) +\dfrac{1}{2}\nu ^3 h \Big(\lambda_1\cos^3\beta  \sin\alpha
	\\ &
	+\lambda_{345}\cos\beta \sin\alpha \sin^2\beta \Big)\bigg)\Bigg]+\rho _2\text{Exp}\Bigg[\Lambda _2\Bigg(-\dfrac{1}{4}\nu ^2 h^2 \Big(\cos^2\alpha  \sin^2\beta  \lambda _2+\lambda_{345} \cos^2\alpha \cos^2\beta \Big) 
	\\ &  
	-\dfrac{1}{2}\nu ^3 h \Big(\lambda _2\cos\alpha \sin^3\beta+\lambda_{345}\cos\alpha \cos^2\beta \sin\beta\Big)\Bigg)\Bigg]+\rho _3\text{Exp}\Bigg[\dfrac{1}{8}\Lambda _3\lambda_1\Big(h^4 \sin^4\alpha
	\\ &
	-4 \nu \cos\beta \sin^3\alpha h^3+6\nu^2 \cos^2\beta\sin ^2\alpha h^2-4\nu ^3 \cos^3\beta \sin\alpha h\Big)\Bigg]+\rho _4\text{Exp}\Bigg[\dfrac{1}{8}\Lambda _4\lambda _2
	\\ &
	\bigg(h^4\cos^4\alpha +4 \nu  \cos ^3\alpha\sin\beta h^3 + 6 \nu ^2 \cos^2\alpha
	\sin^2\beta  h^2+ 4 \nu ^3 \cos\alpha \sin ^3\beta h\bigg)\Bigg] 
	\\ & 
	+\dfrac{1}{4}\lambda_{345} \cos^2\alpha\sin^2\alpha h^4+\dfrac{1}{2} \nu \lambda_{345}  \cos\alpha \sin\alpha\big(\sin\alpha \sin\beta-\cos\alpha \cos\beta\big) h^3 
	\\ &
	+\dfrac{1}{4}\nu^2\lambda_{345}\bigg(\cos^2\alpha \cos^2\beta -4\cos\alpha \cos\beta \sin\alpha \sin\beta+\sin^2\alpha \sin^2\beta \bigg)h^2
	\\ &
	+\dfrac{1}{2}\nu^3\lambda_{345}\cos\beta\sin\beta\big(\cos\alpha\cos\beta-\sin\alpha\sin\beta\big)h +\rho _3\text{Exp}\Bigg[\dfrac{1}{4}\lambda_1\Lambda _3\Big(\sin^2\alpha \sin ^2\beta h^2
	\\ & 
	-2\nu \cos\beta \sin\alpha \sin^2\beta  h\Big)\Big(A^2 +2 {H^C}^2\Big)\Bigg]+\rho _4\text{Exp}\Bigg[ \dfrac{1}{4}\lambda _2\Lambda _4\Big(\cos^2\alpha  \cos ^2\beta h^2
	\\ &
	+2\nu\cos\alpha \cos^2\beta\sin\beta h\Big)\Big(A^2+2 {H^C}^2\Big)\Bigg]+h^2 \Bigg({H^C}^2 \bigg(\dfrac{1}{2}\lambda _3\Big(\cos^2\beta  \sin^2\alpha +\cos^2\alpha  \sin^2\beta \Big)
	\\ &
	+\cos\alpha \cos\beta \sin\alpha \sin\beta\big(\lambda _4+\lambda_5\big)\bigg)+A^2 \bigg(\dfrac{1}{4}\lambda_{34-5}\left(\cos^2\beta  \sin^2\alpha +\cos^2\alpha
	\sin^2\beta \right)
	\\ & 
	+\lambda_5\cos\alpha \cos\beta \sin\alpha \sin\beta\bigg)\Bigg)
	+h \Bigg(\nu  {H^C}^2 \bigg(\lambda _3\big(\cos\alpha \sin^3\beta-\cos^3\beta \sin\alpha \big)+\big(\lambda _4+\lambda_5\big)
	\\ &
	\cos\beta\sin\beta\big(-\cos\alpha \cos\beta+\sin\alpha \sin\beta\big)\bigg)
	+\dfrac{1}{2} \nu  A^2 \bigg(\lambda_{345}\Big(-\cos^3\beta  \sin\alpha+\cos\alpha \sin^3\beta \Big)
	\\ & 
	+2\text{cos$\beta $sin$\beta \lambda $}_5(-\cos\alpha \cos\beta+\sin\alpha \sin\beta)\bigg)\Bigg)
	\end{aligned}
	\right\}
	\qquad \text{}
	\end{equation}
}
\bibliographystyle{ws-ijmpd}
\nocite{*}
\bibliography{Muhammad-Usman.bib}

\end{document}